\documentclass[12pt]{article}
\usepackage{graphicx}

\newcommand\pubnumber{DPF2013-84}
\newcommand\pubdate{\today}

\def\calpoly{Department of Physics, Cal Poly, San Luis Obispo, CA 93407}
\def\fresno{Physics Department, California State University Fresno, 2345 East San Ramon Avenue M/S 37, 
Fresno, California 93740-8031}
\def\bandung{Department of Physics, Faculty of Mathematics and Natural Sciences, 
Institut Teknologi Bandung, Jalan Ganesha 10 Bandung 40132, Indonesia}
\def\emailpj{\footnote{pjones02@calpoly.edu}}
\def\emailds{\footnote{dougs@csufresno.edu}}
\def\emailgm{\footnote{gerardom@csufresno.edu}}
\def\emailtri{\footnote{triyanta@fi.itb.ac.id}}

\def\Title#1{\begin{center} {\Large #1 } \end{center}}
\def\Author#1{\begin{center}{ \sc #1} \end{center}}
\def\Address#1{\begin{center}{ \it #1} \end{center}}

\newcommand\pubblock{\rightline{\begin{tabular}{l} \pubnumber\\
         \pubdate  \end{tabular}}}
\newenvironment{Abstract}{\begin{quotation}  }{\end{quotation}}
\newenvironment{Presented}{\begin{quotation} \begin{center} 
             PRESENTED AT\end{center}\bigskip 
      \begin{center}\begin{large}}{\end{large}\end{center} \end{quotation}}

\def\beq{\begin{equation}}
\def\eeq#1{\label{#1}\end{equation}}
\def\eeqn{\end{equation}}

\def\beqa{\begin{eqnarray}}
\def\eeqa#1{\label{#1}\end{eqnarray}}
\def\eeqan{\end{eqnarray}}

\let\bar=\overbar

\def\Dslash{\not{\hbox{\kern-4pt $D$}}}
\def\dslash{\not{\hbox{\kern-2pt $\del$}}}

\def\msb{{\bar{\ssstyle M \kern -1pt S}}}

\begin{document}
\begin{titlepage}
\pubblock

\vfill
\Title{Field localization and mass generation in an alternative 5-dimensional brane model}
\vfill
\Author{ Preston Jones\emailpj}
\Address{\calpoly}
\Author{ Doug Singleton\emailds \enskip and Gerardo Mu{\~n}oz\emailgm}
\Address{\fresno}
\Author{ Triyanta\emailtri}
\Address{\bandung}\vfill
\begin{Abstract}
This proceedings is from a talk given at the APS DPF 2013 on a 5-dimensional brane world model. This alternative brane world model is formally related but physically distinct from the Randall-Sundrum brane world model. The spin dependent localization of 5D fields for the alternative model are different and in some ways superior to the Randall-Sundrum. The alternative model also exhibits a cutoff in the localization of massive scalar fields not seen in the Randall-Sundrum model. This revision includes a correction to the integrand for the scalar field action appearing in the principle reference \cite{Jones13}.
\end{Abstract}
\vfill
\begin{Presented}
DPF 2013\\
The Meeting of the American Physical Society\\
Division of Particles and Fields\\
Santa Cruz, California, August 13--17, 2013\\
\end{Presented}
\vfill
\end{titlepage}
\def\thefootnote{\fnsymbol{footnote}}
\setcounter{footnote}{0}

\section{Introduction}

Recently, the authors \cite{Jones13} proposed an alternative to the Randall-Sundrum brane world model \cite{RaSu1,RaSu2} (RS model) that is formally similar to the original but has different physical properties. This alternative model does not require fine tuning of the cosmological constant and has a constant energy-momentum in the bulk instead of a vacuum. The alternative model will be identified as the r-metric model following the coordinate of the bulk dimension. This r-metric model shares all the important features of the RS model and in particular the model includes a 4D brane as a topological defect in a uniform 5D bulk. The r-metric model also shares with the RS model spin dependent (spin \textemdash  \enskip  0, 1/2, 1) gravitational confinement of particles, a large extra dimension, 5D and 4D scale proportionality independent of the size of the extra dimension, and constant energy-momentum tensor.

\section{Formal similarity to the Randall-Sundrum model}

The formal similarity between the RS model and r-metric model is apparent by considering the two metrics together. The RS metric with the commonly used $y$ and $z$ coordinate systems is 

\begin{equation}
\begin{array}{*{20}c}
   {ds^2  = e^{ - 2k\left| y \right|} \eta _{\mu \nu } dx^\mu  dx^\nu   - dy^2 ,}  \\
   {\begin{array}{*{20}c}
   {ds^2  = e^{ - 2A\left( z \right)} \left( {\eta _{\mu \nu } dx^\mu  dx^\nu   - dz^2 } \right),} & {e^{ - A\left( z \right)}  = \frac{1}{{1 - 2k\left| z \right|}}},  \\
\end{array}}  \\
\end{array}
\end{equation}

\noindent and the r-metric is

\begin{equation}
\label{rmetric}
\begin{array}{*{20}c}
\label{rmetric}
   {ds^2  = e^{ - 2A\left( r \right)} \left( {\eta _{\mu \nu } dX^\mu  dX^\nu   - dr^2 } \right)},  \\
   {A\left( r \right) = k\left| r \right|}.  \\
\end{array}
\end{equation}

\noindent Both metrics are conformally flat and transformations can be readily identified between all three sets of coordinates, with the RS metric most obviously conformally flat in the $z$ coordinates. The RS metric in the $z$ coordinates is sufficiently similar to the r-metric that some care must be exercised to distinguish between the two. The transformation between the RS metric and r-metric ${dX^\mu   = \frac{{e^{ - k\left| y \right|} }}{{1 - k\left| y \right|}}dx^\mu  }$ and ${e^{ - k\left| r \right|}  = 1 - k\left| y \right|}$ is not an exact differential, ${dX^\mu   = A\left( {x^\mu  ,y} \right)dx^\mu   + B\left( {x^\mu  ,y} \right)dy}$, since ${\partial _y A \ne 0}$ and the $y$ and $r$ coordinates do not represent the same geometry. However, on any foliation $y=constant$ the two geometries are connected by a simple coordinate transformation.
 
We will now show how both models can be considered topological defects in a 5D bulk. Solving the Einstein-Hilbert equation, $ G_{AB}  + g_{AB}  = \kappa ^2 T_{AB} $, the energy-momentum tensor for the RS model is

\begin{equation}
\begin{array}{*{20}c}
   {\eta _{\mu \nu } e^{ - 2k\left| y \right|} \left( {6k^2  + \lambda _{[y]} } \right) - 6k\eta _{\mu \nu } \delta \left( y \right) = \kappa ^2 T_{\mu \nu } }  \\
   { - \left( {6k^2  + \lambda _{[y]} } \right) = \kappa ^2 T_{55} }  \\
\end{array}
\end{equation}

\noindent and for the r-metric

\begin{equation}
\begin{array}{*{20}c}
   {\eta _{\mu \nu } \left( {3k^2  + e^{ - 2k\left| r \right|} \lambda _{[r]} } \right) - 6k\eta _{\mu \nu } \delta \left( r \right) = \kappa ^2 T_{\mu \nu } }  \\
   { - 6k^2  - e^{ - 2k\left| r \right|} \lambda _{[r]}  = \kappa ^2 T_{55} }.  \\
\end{array}
\end{equation}

\noindent Requiring that the RS model be consistent with a 4D brane that is a topological defect in a uniform 5D bulk the cosmological constant must be fine tuned, $ \lambda _{[y]} = -6k^2 $. For the r-metric model to be consistent with a 4D brane as a topological defect in a uniform 5D bulk the cosmological constant must vanish $ \lambda _{[r]} = 0 $. 

\section{Spin dependent localization of fields to the brane}

The condition for the localization of fields to the brane is that the Wick rotated propagator does not vanish, $\int {Dx} \,e^{ - S}  \to \;S = finite$. Considering the action for spinor fields $ S_\psi   = \int {d^5 x\sqrt g } \bar \Psi i\Gamma ^M D_M \Psi $ and expanding the integral produces three terms. The two kinetic terms are

\begin{equation}
\begin{array}{*{20}c}
   {c^2 \int\limits_0^\infty  {dr} \,e^{ - 2mr} \int {d^4 x} \,\bar \psi _R i\gamma ^\mu  \partial _\mu  \psi _R  \to finite}  \\
   {d^2 \int\limits_0^\infty  {dr} \,e^{2mr} \int {d^4 x} \,\bar \psi _L i\gamma ^\mu  \partial _\mu  \psi _L  \to \infty }  \\
\end{array}
\end{equation}

\noindent and a gamma 5 term is
 
\begin{equation}
c\,d\int\limits_0^\infty  {dr} \,\int {d^4 x} \,\left( {\psi ^\dag  _L \psi _R  + \psi ^\dag  _R \psi _L } \right) \to \infty 
\end{equation}

\noindent where $c$ and $d$ are constants. Since the action for the spinor field is not finite there can be no localized spinor fields due to gravitational confinement. Localization of spinor fields to the brane would require a mechanism other than gravitational confinement which is also the case for the RS model with decreasing warping as shown in Figure~\ref{fig:spins}.

The action for the massless gauge fields is finite for decreasing warping as can be seen by expanding the action with the r-metric (\ref{rmetric}),
\begin{equation}
\begin{array}{*{20}c}
   {S_A  =  - \frac{1}{4}\int {d^5 x} \sqrt g g^{MN} g^{RS} \left( {\partial _M A_N  - \partial _N A_M } \right)\left( {\partial _R A_S  - \partial _S A_R } \right)}  \\
   { =  - \frac{{c^2 }}{4}\int\limits_0^\infty  {dr} \,e^{ - kr} \int {dx^4 } \,\left( {\partial ^\mu  a_\mu   - \partial ^\nu  a_\nu  } \right)\left( {\partial ^\sigma  a_\sigma   - \partial ^\rho  a_\rho  } \right)},  \\
\end{array}
\end{equation}

\noindent where $ {A_r  = constant}$ and ${A_\mu  \left( {x^M } \right) = a_\mu  \left( {x^\nu  } \right)c\left( r \right)} $. The finite action for gauge fields localizes the massless spin 1 particles to the brane. This gravitational confinement of massless gauge fields is an improvement over the RS model as shown in Figure~\ref{fig:spins}.

%%%%%%%%%%%%%%%%%%%%%%%%%%%%%%%%%%%%%%%%%%%%%%%%%%%%

\begin{figure}[htb]
\centering
\[
\begin{array}{*{20}c}
   {Spin} & {RS\;model} & {r - metric}  \\
   0 & { - 2k} & {\begin{array}{*{20}c}
   { - 2k \to 2m < 3 k}  \\
   { + 2k \to 2m < 3k}  \\
\end{array}}  \\
   1 & {none} & { - 2k}  \\
   {{\textstyle{1 \over 2}}} & { + 2k} & {none}  \\
\end{array}
\]
\caption{Comparison of spin dependent localization for the Randall-Sundrum and r-metric models. The negative values $-2k$ are for decreasing warping and positive values $+2k$ are for increasing warping.}
\label{fig:spins}
\end{figure}

%%%%%%%%%%%%%%%%%%%%%%%%%%%%%%%%%%%%%%%%%%%%%%%%%%%%

The action for scalar fields in the r-metric model using equation (\ref{rmetric}) is
\begin{equation}
\label{baction}
\begin{array}{*{20}c}
   {S_\Phi   = \int {d^5 x} \sqrt g g^{MN} \partial _M \Phi ^* \partial _N \Phi  = N\left( {g^{\mu \nu } } \right)\int {d^4 x\,} \partial _\mu  \phi \,\partial _\nu  \phi  + M^{2}_{0} \int {d^4 x\,} \phi ^2 }  \\
   {N = \int\limits_0^\infty  {dr} \sqrt g \chi ^2 g^{\mu \nu }  = \left\{ {\begin{array}{*{20}c}
   {1\quad if\quad 3\left| k \right| > 2m}  \\
   {\infty \quad if\quad 3\left| k \right| < 2m}  \\
\end{array}} \right.}  \\
   {M^{2}_{0}  = \int\limits_0^\infty  {dr} \sqrt g g^{rr} \partial _r \chi ^* \partial _r \chi  = \left\{ {\begin{array}{*{20}c}
   {M^{2}_{0}\left( {m,k} \right)\;\;if\; 3 \left| k \right| > 2m}  \\
   {\infty \quad if\quad 3 \left| k \right| < 2m}  \\
\end{array}} \right.} \\
\end{array}
\end{equation}

\noindent where $\Phi  = \phi \left( {x^\mu  } \right)\chi \left( r \right)$ and

\begin{equation}
\label{masses}
M_0\left( {m,k} \right)  = \frac{3}{2}k -  \frac{1}{2} \sqrt {9k^2  - 4m^2 }  .
\end{equation}

\noindent The action is finite and the field localized to the brane for masses below $m < \frac{3}{2}k$ with decreasing warping and $m < \frac{{3 }}{2}k$ for increasing warping as shown in reference \cite{Jones13}. The comparison of the localization of scalar fields in the r-metric and Randall-Sundrum models is provided in Figure~\ref{fig:spins} which includes both decreasing and increasing warping. Unlike the RS model with decreasing warping the r-metric model exhibits a cutoff in the mass spectra localized to the brane.

%\section{Prediction of two massive scalar boson}

%The schematic relation between the two masses and the mass cutoff are shown in Figure~\ref{fig:mass} with the mass cutoff at $k  \frac{{\sqrt 5 }}{2}$ for the decreasing warping.

Setting the mass in the action for the scalar boson in equation (\ref{masses}) equal to the mass in the 4D equation of motion $M_0 \left( {m,k} \right) = m$ has the solution, $ m_{\left(  \pm  \right)}^2 =  0$. This is in contrast to the conclusion reached earlier in reference \cite{Jones13} due to an error in the integrand for the scalar field action equation (30). This leaves the parameters $m$ and $k$ undetermined.  One way to establish the magnitude of $k$ is in the reduction of the 5D action to an effective 4D action \cite{RaSu2,benson},

\begin{equation}
\label{k}
\begin{array}{*{20}c}
   {S_{gravity}  =  - 2M^3 \int {dx^4 \int\limits_{ - \infty }^\infty  {dr} } \sqrt g R =  - 2M^3 \left( {\frac{2}{{3k}}} \right)\int {dx^4 } \sqrt {{}^{\left( 4 \right)}g} \;{}^{\left( 4 \right)}R}  \\
   { \to M_{Pl}^2  = M^3 \left( {\frac{2}{{3k}}} \right)}  \\
\end{array}
\end{equation}

\noindent where $ \sqrt g=e^{-5k \left| r \right|} $ and $ R = e^{2k\left| r \right|} \eta ^{\mu \nu } R_{\mu \nu } $ and integrating over an infinite extra dimension. The bulk warping is  $k = \frac{{2M^3 }}{{3M_{Pl}^2 }}$ which is proportional to the ratio between the 5D and 4D scales.

%%%%%%%%%%%%%%%%%%%%%%%%%%%%%%%%%%%%%%%%%%%%%%%%%%%%%%%%%%%%%%%%%%%%%%%%%
%%%%%%%%%%%%%%%%%%%%%%%%%%%%%%%%%%%%%%%%%%%%%%%%%%%%%%%%%%%%%%%%%%%%%%%%%
%%
%%   use this format to include an .pdf figure into your paper
%%

%\begin{figure}[htb]
%\centering
%\includegraphics[height=2.75in]{mass}
%\caption{Schematic relation of the two massive scalar bosons and the mass cutoffs. The upper cutoff ($k\frac{3}{2}$) is for the increasing warping and the lower ($k\frac{\sqrt{5}}{2}$) for the decreasing. For increasing warping both bosons $m_{\left(  +  \right)} $ and $m_{\left(  -  \right)} $ are localized and for decreasing warping only the lower mass boson $m_{\left(  -  \right)} $ is localized.}
%\label{fig:mass}
%\end{figure}

%%%%%%%%%%%%%%%%%%%%%%%%%%%%%%%%%%%%%%%%%%%%%%%%%%%%

% as shown in Figure~\ref{fig:mass}.

%\section{Discussion and conclusions}

%In conclusion, ...

%\Acknowledgments

\end{document}